\documentclass[runningheads]{llncs}
\usepackage[T1]{fontenc}
\usepackage{graphicx}
\usepackage{cite}
\usepackage{amsmath,amssymb,amsfonts}
\usepackage{algorithmic}
\usepackage{graphicx}
\usepackage{textcomp}
\usepackage{xcolor}
\usepackage{hyperref}       
\usepackage{url}            
\usepackage{booktabs}       
\usepackage{amsfonts}       
\usepackage{nicefrac}       
\usepackage{microtype}      
\usepackage{subcaption}
\usepackage{float}
\usepackage{tabularx}
\usepackage{ctable}
\usepackage{caption}
\usepackage{adjustbox}

\definecolor{darkred}{rgb}{1.0, 0.0, 0.0}
\begin{document}
\title{LLM-Enhanced Feature Engineering for Multi-Factor Electricity Price Predictions}
%
%
\author{Haochen Xue \inst{1,\footnotemark[1]}\and Chenghao Liu\inst{3,\footnotemark[1]} \and
Chong Zhang \inst{1,\footnotemark[1]} \and Yuxuan Chen\inst{2}  \and \\ Angxiao Zong \inst{1}  \and  Zhaodong Wu \inst{1} \and Yulong Li  \inst{1}  \and Jiayi Liu \inst{1}  \and  Kaiyu Liang \inst{1} \and \\    Zhixiang Lu \inst{1} \and Ruobing Li \inst{1}    \and Jionglong Su\inst{1\dagger} }

\authorrunning{H Xue et al.}
\institute{School of AI and Advanced Computing, Xi'an Jiaotong-Liverpool University \and
Columbia University  \textsuperscript{3} Tongji University
}
\maketitle 
\def \thefootnote{${*}$}\footnotetext{Equal contribution.\\
\hspace*{-0.85em}{$\dag$} Corresponding email: Jionglong.Su@xjtlu.edu.cn}
\begin{abstract}
Accurately forecasting electricity price volatility is crucial for effective risk management and decision-making. Traditional forecasting models often fall short in capturing the complex, non-linear dynamics of electricity markets, particularly when external factors like weather conditions and market volatility are involved. These limitations hinder their ability to provide reliable predictions in markets with high volatility, such as the New South Wales (NSW) electricity market. To address these challenges, we introduce \textbf{FAEP}, a \textbf{F}eature-\textbf{A}ugmented \textbf{E}lectricity Price \textbf{P}rediction framework, FAEP leverages Large Language Models (LLMs) combined with advanced feature engineering to enhance prediction accuracy. By incorporating external features such as weather data and price volatility jumps, and utilizing Retrieval-Augmented Generation (RAG) for effective feature extraction, FAEP overcomes the shortcomings of traditional approaches. A hybrid XGBoost-LSTM model in FAEP further refines these augmented features, resulting in a more robust prediction framework. Experimental results demonstrate that FAEP achieves state-of-art (SOTA) performance compared to other electricity price prediction models in the Australian New South Wale electricity market, showcasing the efficiency of LLM-enhanced feature engineering and hybrid machine learning architectures.
\end{abstract}

\section{Introduction}

Electricity is a fundamental resource for modern societies, and its price fluctuations have a profound impact on economic stability and energy management~\cite{heydari2020short}. Predicting these fluctuations accurately is crucial for ensuring sustainable economic growth and social welfare~\cite{mohammadi2012application}. However, electricity price forecasting is particularly challenging due to the complexity and unpredictability of electricity markets, which are subject to a wide range of factors, including non-stationarity, multi-seasonality, and external market influences~\cite{shao2021feature}. These challenges are especially pronounced in markets like the Australian New South Wales (NSW) electricity market, which is one of the most liberalized and free electricity markets in the world. The freedom of the market leads to significant price volatility. Price fluctuations are driven by factors such as unpredictable weather patterns, sudden supply-demand imbalances, and frequent regulatory changes~\cite{rai2020australia}. The inherent unpredictability and dynamic nature of these elements make forecasting in such markets incredibly challenging, as they are subject to rapid shifts that can be difficult to anticipate with precision. 

Existing forecasting methods for the NSW electricity market include traditional statistical methods like ARIMA~\cite{box2015time} and GARCH~\cite{laurent2012forecasting}, which struggle to capture the non-linearity and complex dynamics of electricity prices. Deep learning methods such as LSTM~\cite{xue2024multi}, Temporal Convolutional Networks~\cite{lara2020temporal}, and Temporal Fusion Transformers~\cite{lim2021temporal} are able to capture the non-linearity and complex dynamics of electricity prices to improve prediction accuracy. However, they often fail to properly account for external factors, such as weather conditions and market dynamics, which significantly influence electricity price fluctuations. Though they can include external factors as inputs, these models lack the ability to assess their true relevance or relationship with electricity prices at different time steps, leading to increased model dimensionality and the introduction of irrelevant noise, which degrades prediction performance.

To overcome these limitations, we introduce \textbf{FAEP}, a framework that leverages the capabilities of LLMs for advanced feature engineering. This framework consists of the following two main components: \textbf{(i) LLM-based Feature Augmentation and Selection:} This component focuses on utilizing LLMs to enhance and select the most relevant features for the prediction task. \textbf{(ii) Hybrid Model:} This part employs a hybrid model combining LSTM and XGBoost, leveraging the selected features for improved prediction accuracy.

\textit{LLM-based Feature Augmentation and Selection:} Direct application of LLMs in feature engineering often leads to issues such as hallucinations and inaccurate responses. To mitigate this, we utilize Retrieval-Augmented Generation (RAG)~\cite{lewis2020retrieval} to incorporate annual weather report data from NSW during 2009 to 2019. This external data enables LLMs to address domain-specific questions beyond their internal knowledge base, providing more accurate features. Our feature engineering process incorporates temperature, historical prices, and market dynamics to select the most relevant features. We use a combination of forward sequential selection and Kernel Principal Component Analysis (KPCA)~\cite{du2021deep} to optimize feature dimensionality, which boosts model performance. Additionally, we classify volatility factors into short-term, medium-term, and long-term categories, based on the impact of regional market demand and price fluctuations. This classification allows the model to account for the differing effects of these volatility factors across various time frames, improving its overall predictive performance. Furthermore, we introduce novel features like continuity of price transitions (CV) and price jumps (PJ) to capture both continuous and jump components of realized volatility. These selected features significantly improve the predictive accuracy of FAEP. In the ablation study, the model with these features showed an improvement of approximately 64.35\% compared to the model without them.


\textit{Hybrid Model:} In FAEP, we propose a hybrid XGBoost-LSTM model designed to improve both accuracy and performance in predicting electricity price fluctuations. The XGBoost~\cite{wang2020forecasting} component is well-suited for handling high-dimensional data and feature selection, capturing nonlinear relationships in the input data. The LSTM~\cite{song2020time} component, on the other hand, excels at modeling long- and short-term dependencies in time series data. By combining the strengths of both models, our approach is better equipped to handle the complexities of electricity price volatility. Additionally, to estimate the overall price variation and investigate the characteristics of electricity price volatility, we incorporate the Realized Volatility (RV) method~\cite{bucci2017forecasting}. Given electricity's non-storability and short-term demand's inelasticity, electricity prices exhibit significant volatility with pronounced multi-seasonal effects. By integrating RV, FAEP can more accurately capture these dynamics, leading to 43.23\% improvement in predictions. 

\noindent The main contributions of this study are as follows: 
\vspace{-0.2cm}
\begin{itemize}
    \item We propose an innovative feature engineering framework based on RAG-LLM, which improves both prediction accuracy and interpretability.
    \item Building on enhanced features, we design a hybrid XGBoost-LSTM and LLM model to handle high-dimensional data, feature selection, and model both long- and short-term dependencies in time series.
    \item Our experiments demonstrate FAEP's SOTA performance over other frameworks, leading to improvements of average 48.67\% improvement in matrices like MAE, MSE, and MAPE compared to other electricity forecasting frameworks.
\end{itemize}

\section{Related Work}

\subsection{Time Series Forecasting}

Time series forecasting has progressed significantly from traditional methods like ARIMA~\cite{box2015time} and GRACH~\cite{laurent2012forecasting}, which are limited to modeling linear patterns and cannot handle non-linear relationships in the data. Building on these limitations, more complex machine learning methods have been introduced. Recurrent Neural Networks (RNNs)~\cite{schmidt2019recurrent} and Long Short-Term Memory (LSTM) networks \cite{song2020time} are able to model temporal dependencies effectively, making them suitable for sequential data. Recent advancements include attention-based models like Temporal Fusion Transformers (TFT) \cite{lim2021temporal} and Temporal Convolutional Networks (TCNs) \cite{bai2018empirical}, which offer improved performance in capturing long-range dependencies and complex patterns within time series data. Additionally, hybrid methods such as Facebook's Prophet combine statistical models with machine learning techniques to enhance forecasting accuracy \cite{taylor2018forecasting, liu2025incompletemodalitydisentangledrepresentation}. However, these methods overlook the impact of external factors on electricity prices, focusing solely on historical price data, which leads to inaccurate predictions \cite{10.1145/3715073.3715083, liu2024mtsasnnmultimodaltimeseries}.

\subsection{Retrieval Augmented Generation}

Retrieval Augmented Generation (RAG) enhances natural language processing by combining retrieval-based and generation-based methods~\cite{salemi2024evaluating, jin2024prollm}. Traditional generative models often struggle with factual accuracy, relying solely on internal knowledge~\cite{chen2023beyond, 10.1007/978-3-031-78183-4_30}. RAG addresses this by integrating an external retrieval mechanism to fetch relevant information, improving context and accuracy \cite{lewis2020retrieval,jin2025two}. This approach has shown significant improvements in tasks like question answering and dialogue systems by reducing hallucinations and enhancing coherence \cite{zhang2024target,xue2025mmrc,10884369, tang2025intervening,zhou2025multimodalsituationalsafety, 10.1007/978-3-031-72347-6_1, 10.1007/978-981-96-2681-6_23,zhu2025llm}. 

\section{FAEP}
\subsection{Problem statement}
\begin{figure*}[!h]
\centering
\vspace{-10pt}
\includegraphics[width=1.0\textwidth]{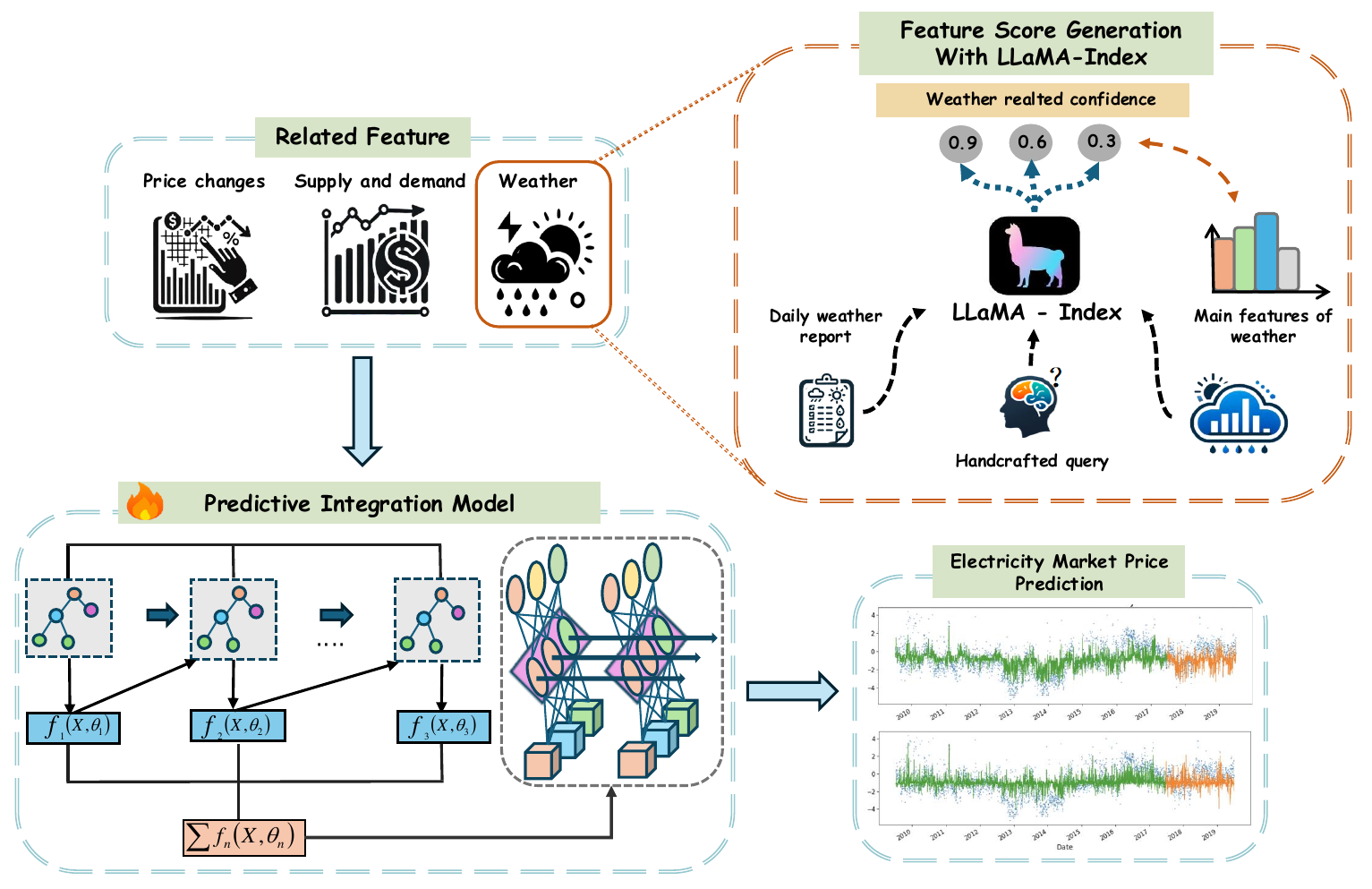} 
\caption{The Overview of the LLM Integrated XGBoost-LSTM Model, demonstrating the process of predicting electricity market prices by evaluating and integrating relevant features.}
\vspace{-25pt}
\label{fig:xgboost-lstm}
\end{figure*}

We aim to forecast daily electricity price volatility, denoted as day $t \in \mathcal{X}$. Prices are observed at discrete time points $j = 1, 2, ..., M$ within each day. Let's$r_{i,j}$ represent the price variation at the $j$-th time point on day $i$, defined as:
\begin{equation}
    r_{i,j} = p_{i,j} - p_{i,j-1}, \quad RV_t = \sum_{j=1}^{M}r^2_{i,j},
\end{equation}

where \( r_{i,j} \) is the price difference between adjacent time points, and the realized volatility \( RV_t \) is the sum of the squared variations \( r^2_{i,j} \) for day \( t \). Traditional regression models often struggle with accurately forecasting volatility with many external factors~\cite{yuan2025machine}. To address this, we adopt an ensemble approach that combines LLMs with the XGBoost-LSTM model. This hybrid model structure follows a two-step training and prediction process, refining residuals to enhance forecast accuracy. The LLM improves feature analysis, better managing the impacts and interactions of multiple factors. As discussed in Section \ref{sec: XGBLSTM}, this approach significantly increases the accuracy and reliability of volatility forecasts. Fig. \ref{fig:xgboost-lstm} illustrates the structure of our framework. FAEP predicts the process of electricity market prices by evaluating and integrating relevant features such as price changes, supply and demand, and weather. What is locally reflected is our framework for forecasting electricity prices by using the Predictive Integration Model and RAG mainstream Llama-Index model.

\subsection{Feature Selection Mechanism}
\subsubsection{Feature Selection:}

Feature engineering is crucial in electricity market forecasting and directly affects model performance. To achieve effective feature selection, we examine three key aspects: weather, the balance between supply and demand, and price changes \cite{lago2018forecasting, wen2025feature}. Based on them, we identify corresponding key features, including supply and demand status, retail prices, and related variables.

We additionally select two key features to predict electricity price fluctuations: the continuity of price transitions within a day ($CV_t$) and significant price jumps ($PJ_t$). Discontinuous market events (jumps) may cause notable price shifts, which are detected using the statistical metric $Z_t$, calculated as follows:
\begin{equation}
    Z_t = \frac{RV_t - BPV_t}{\sqrt{2\left(\frac{\pi}{2}\right)^2 + \frac{\pi^{-5}}{M}\max\left(1, \frac{BPV_t}{TPQt}\right)^2}},
\end{equation}

\noindent where $RV_t$ is the realized volatility, $BPV_t$ is the bipower volatility extracted from the dataset denoting uplink and downlink grid electricity prices, and $TPQ_t$ is the tri-power quarticity used to estimate integrated variance \cite{barndorff2004power}.
$PJ_t$ is a key evaluation metric representing significant price jumps, the price jump in electricity price refers to the sharp fluctuations or sudden changes in the spot price in a short time in the electricity market. This metric will then be compared with the predefined threshold $\phi_a$, calculated as follows:
\begin{equation}
 J_t = I(Z_t > \phi_a) \cdot (RV_t - BPV_t)
\vspace{-10pt}
\end{equation}

\subsubsection{Dimensionality Reduction and Feature Extraction:}

To reduce feature complexity and mitigate the risk of overfitting, we first apply Sequential Feature Selection (SFS) to identify an optimal subset of 30 key features. This process enhances the performance of the XGBoost-LSTM model. Additionally, we employ Kernel Principal Component Analysis (KPCA) for dimensionality reduction, which further improves the model's accuracy in predicting price fluctuations.

\subsection{Hybrid XGBoost-LSTM Network}
\label{sec: XGBLSTM}
The integrated XGBoost-LSTM model adopts a two-stage ensemble approach to enhance predictive performance by combining the strengths of both models. In the first stage, XGBoost captures complex feature relationships and nonlinear patterns, generating initial predictions. In the second stage, LSTM refines these predictions by effectively handling temporal dependencies. The final prediction is obtained by analyzing the errors from both models, $\epsilon_1$ (LSTM) and $\epsilon_2$ (XGBoost), to calculate weight coefficients $\omega_1$ and $\omega_2$, $\epsilon_1$ and $\epsilon_2$ are the residuals of the LSTM and XGBoost models, respectively. These weight coefficients are then used to combine the predictions $P_{RV_1t}$ and $P_{RV_2t}$ through weighted aggregation:
\begin{equation}
P_{RV} = \omega_1 f_{RV_1t} + \omega_2 f_{RV_2t}, \quad t=1,2,\ldots,n \ ,
\end{equation}
\begin{equation}
\begin{aligned}
\omega_1 &= \frac{\epsilon_2}{\epsilon_1 + \epsilon_2} \qquad & \omega_2 &= \frac{\epsilon_1}{\epsilon_1 + \epsilon_2} \ .
\end{aligned}
\end{equation}
\subsection{Integration of LLM with XGBoost-LSTM}
We integrate LLM with tetrival augmented generation to enhance feature extraction and prediction using weather data from NSW spanning from 2009 to 2019. The process begins by retrieving and analyzing weather reports through LLMs, incorporating a confidence scoring mechanism to evaluate the relevance of weather-related features. These features are then integrated into the XGBoost-LSTM model to improve the prediction of electricity price volatility.
\vspace{-0.4cm}
\subsubsection{Retrieving annual weather report:} We conduct academic searches via Google to obtain NSW's annual weather reports from 2009 to 2019, primarily in PDF format. This data provides valuable external knowledge that supports document embedding and LLM-based retrieval, facilitating the analysis of meteorological trends to improve electricity price volatility predictions.
\vspace{-0.4cm}
\subsubsection{Confidence Score of LLaMA-Index:} Using GPT-3.5-turbo, we analyze the impact of weather features on price volatility. The LLaMA-Index assigns scores (ranging from 1 to 5) to weather-related queries, generating a ‘weather rating’ indicator. This indicator is then incorporated into the XGBoost-LSTM model, leveraging natural language processing (NLP) and LLMs to enhance prediction accuracy and enable more precise forecasting.

\section{Experiments}

\subsection{Data and Statistics}

Our analysis examines high-frequency spot price, demand, and supply data from the Australian National Electricity Market (NEM), which includes New South Wales (NSW) shown in Table~\ref{tab:1}, Victoria (VIC), Queensland (QLD), Tasmania (TAS), and South Australia (SA). Managed by the Australian Energy Market Operator (AEMO), the NEM operates as a real-time spot market, setting electricity prices every five minutes. These prices are averaged to produce half-hourly trading spot prices, which are accessible on AEMO's website \footnote{http://www.aemo.com.au/}. The dataset covers a period from 2 July 2009 to 1 July 2019, spanning 10 years (3653 days), with 48 half-hour intervals per day, resulting in a total of 175,344 observations. Data anomalies are addressed by interpolating non-positive prices using the nearest positive values. Descriptive statistics reveal considerable volatility in electricity prices, marked by significant deviations and high kurtosis. All prices are expressed in Australian Dollars (\$) per megawatt hour.
\begin{table*}[!ht]
    \vspace{-25pt}
    \caption{Statistics for Electricity Prices of the NSW Market}
    \centering
    \small
    \begin{tabularx}{0.8\textwidth}{cccccccccc}
        \toprule
        \#obs & Mean & SD & Skew & Kurt & Min & 25\% & Median & 75\% & Max \\
        \midrule 
        175344 & 43.93 & 203.12 & 38.55 & 1703.81 & 1.45 & 22.93 & 30.47 & 47.84 & 13742.68 \\
        \bottomrule
    \end{tabularx}
    \label{tab:1}
    \vspace{-20pt}
\end{table*}

Electricity prices exhibit unique intraday, intraweek, and seasonal patterns. Following \cite{haugom2012forecasting}, we demean intraday returns using the half-hourly median return $\mu_{t, j}=\bar{r}{mn, dy, hr}$, $\mu_{t, j}$This variable is hourly granularity electricity bill change data. Based on data from 1 July 2009 to 30 June 2017. The demeaned returns $r{t, j}^*=r_{t, j}-\mu_{t, j}$ are used to calculate several realized estimators, and the statistics for the NSW market are shown in Table \ref{tab:2}. With a 99\% jump detection significance level, the mean jump (0.242) accounts for nearly 32\% of daily realized volatility (0.733), highlighting the impact of price jumps on volatility and justifying their inclusion in the prediction model. 
\begin{table*}[ht]
\vspace{-20pt}
\caption{Statistics for the Realized Estimators of the NSW Electricity Prices}
\centering
\small
\begin{tabularx}{0.8\textwidth}{XXXXXXXXXX}
\toprule
\text{} & \text{Mean} & \text{Median} & \text{Min} & \text{Max} & \text{SD} & \text{Skew} & \text{Kurt} \\
\midrule 
RV$_{\text{t}}$ & 1.134 & 0.358 & 0.035 & 124.745 & 3.891 & 13.416 & 257.810  \\
J$_{\text{t}}$ & 0.3681 & 0.000 & 0.000 & 93.187 & 2.741 & 18.392 & 430.136 \\
CV$_{\text{t}}$ & 0.7016 & 0.307 & 0.018 & 81.490 & 2.306 & 17.973 & 441.602 \\
lnRV$_{\text{t}}$ & -1.104 & -1.027 & -3.352 & 4.826 & 1.078 & 1.362 & 7.046 \\
ln(J$_{\text{t}}$+1) & 1.109 & 0.000 & 0.000 & 4.545 & 0.366 & 6.693 & 60.083 \\
lnCV$_{\text{t}}$ & -1.247 & -1.180 & -4.017 & 4.400 & 0.961 & 0.998 & 5.912 \\
$\sqrt{\text{RV}_{\text{t}}}$ & 0.733 & 0.598 & 0.187 & 11.169 & 0.652 & 6.175 & 47.185 \\
$\sqrt{\text{J}_{\text{t}}}$ & 0.242 & 0.000 & 0.000 & 9.653 & 0.537 & 7.356 & 71.049 \\
$\sqrt{\text{CV}_{\text{t}}}$ & 0.631 & 203.12 & 38.55 & 9.027 & 0.464 & 6.148 & 52.741 \\
\bottomrule
\end{tabularx}
\label{tab:2}
\vspace{-30pt}
\end{table*}

\subsection{Forecast comparison results}

\begin{figure*}[!ht]
    \centering
    \vspace{-10pt}
    \includegraphics[width=0.75\textwidth]{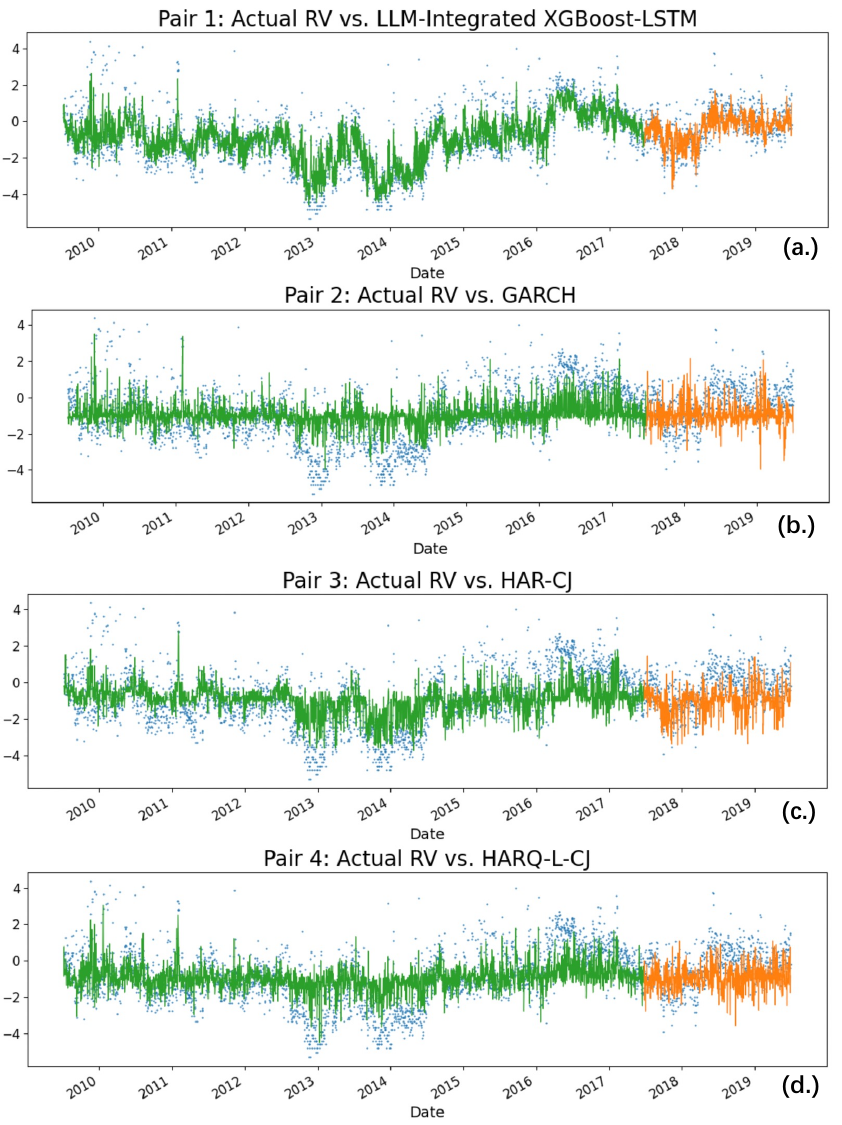}
    \vspace{-10pt}
    \caption{Forecast Comparison Results}
    \label{fig:picture005}
    \vspace{-20pt}
\end{figure*}

We compare the predictive performance of four models: FAEP (Ours), GARCH \cite{laurent2012forecasting}, HAR-CJ \cite{ozbekler2021volatility}, and HARQ-L-CJ \cite{4610781}, using realized volatility as the evaluation metric. These are models specifically designed for electricity price forecasting. The results are visualized in Fig. \ref{fig:picture005}. In each sub-figure, the test data is represented by the orange line, while the training data is shown in green, with the model names indicated in the titles. As seen in Fig. \ref{fig:picture005}(a), the LLM-Integrated XGBoost-LSTM model exhibits only a slight deviation between the predicted and actual values, indicating high accuracy. In contrast, the other models show flatter predictions, reflecting a lower level of accuracy.
To further evaluate the performance of these models, we assessed their Mean Squared Error (MSE) and Mean Absolute Error (MAE), as shown in Table \ref{tab:3}. MSE emphasizes larger prediction errors by squaring them, while MAE provides an average of absolute errors, which is less affected by outliers. The results indicate that our FAEP model significantly outperforms the other models, with MAE and MSE reduced by approximately 32.18\% and 22.21\% compared to the next best-performing model, HARQ-L-CJ, demonstrating its superior prediction accuracy.
\begin{table}[!ht]
\centering
\vspace{-25pt}
\caption{MSE, MAE and MAPE for different models. Figures in \textbf{Bold} denotes the best result for each metric.} 
\label{tab:3}
\small
    \begin{tabular}{lccc}
    \hline 
    \textbf{Model} & \textbf{MAE}$\downarrow$  & \textbf{MSE} $\downarrow$  & \textbf{MAPE}$\downarrow$  \\
    \hline 
    GARCH & 1.70 & 1.32 & 1.21 \\
    HAR-CJ & 1.19 & 1.01 & 0.87 \\
    HARQ-L-CJ & 0.87 & 0.81 & 0.69 \\
    \hline
    \textbf{FAEP(Ours)} & \textbf{0.59} & \textbf{0.63} & \textbf{0.54} \\ 
\hline
\end{tabular}
\end{table}

\begin{figure}[!ht]
    \centering
    \vspace{-5pt}
    \includegraphics[width=0.5\linewidth]{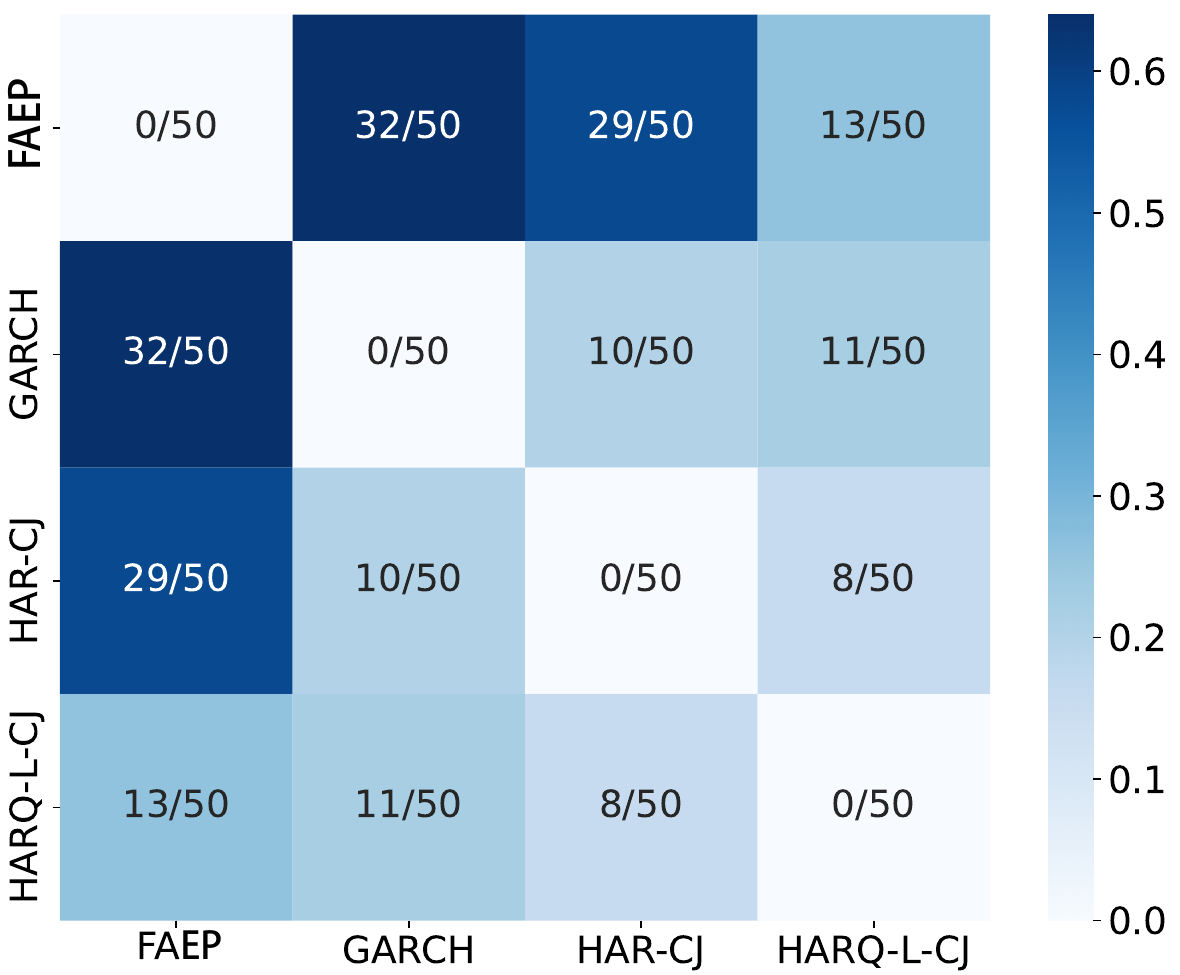}
    \caption{Diebold–Mariano Test Rejection Rates for the Out-of-Sample Volatility Forecasts.} 
    \vspace{-22pt}
\label{fig:heatmap}
\end{figure}

Building on the evaluation metrics, we compared the models using the one-sided Diebold-Mariano (DM) test \cite{diebold2002comparing}, as shown in Fig.~\ref{fig:heatmap}, which presents the DM test rejection rates for our four models during the out-of-sample forecast period (2017–2019). Each segment, consisting of 50 instances, evaluates new forecasts using MSE. The DM test compares model pairs under the null hypothesis of "equal forecast accuracy" against the alternative hypothesis of "significant differences," with a 10\% significance level. The heatmap reveals significant disparities, where, for example, an entry of 32/50 indicates that the DM test rejected the null hypothesis of equal accuracy between the FAEP and GARCH models in 32 out of 50 instances. Notably, higher rejection rates in the first column suggest that the FAEP shows more distinct differences than the other models. With its smaller MSE, the FAEP model demonstrates superior forecasting performance, enhancing out-of-sample accuracy.

\subsection{Ablation Study}

To assess FAEP's performance accurately, we employ two key evaluation strategies: deconstruction analysis of the model's core components and feature selection.

\begin{table*}[h]
    \centering
    \footnotesize
    \vspace{-20pt}
    \caption{Ablation Study of Model Components and Feature Selections, 'w/o' means without. Figures in \textbf{Bold} denotes the best result for each matrix.}
    \begin{tabular}{lcccc}
        \hline 
        \textbf{Configuration}  & \textbf{MAE}$\downarrow$ & \textbf{MSE}$\downarrow$ & \textbf{MAPE}$\downarrow$ \\
        \hline 
        FAEP w/o XGBoost & 1.35 \textcolor{darkred}{\scriptsize$\downarrow0.76$} & 1.57\textcolor{darkred}{\scriptsize$\downarrow0.94$} & 1.28\textcolor{darkred}{\scriptsize$\downarrow0.74$}\\
        FAEP w/o LSTM &   0.77 \textcolor{darkred}{\scriptsize$\downarrow0.18$} & 1.32 \textcolor{darkred}{\scriptsize$\downarrow0.68$} & 0.73 \textcolor{darkred}{\scriptsize$\downarrow0.19$}\\
        FAEP w/o LLM-Augmented Feature &  1.63 \textcolor{darkred}{\scriptsize$\downarrow1.04$} & 1.79 \textcolor{darkred}{\scriptsize$\downarrow1.16$} & 1.6 \textcolor{darkred}{\scriptsize$\downarrow1.06$}\\
        \hline
        Feature w/o Weather & 1.85 \textcolor{darkred}{\scriptsize$\downarrow1.26$} & 1.92 \textcolor{darkred}{\scriptsize$\downarrow1.29$} & 1.76 \textcolor{darkred}{\scriptsize$\downarrow1.22$}\\
        Feature  w/o  Market Supply and Demand &  1.75 \textcolor{darkred}{\scriptsize$\downarrow1.16$} & 1.83 \textcolor{darkred}{\scriptsize$\downarrow1.20$} & 1.66 \textcolor{darkred}{\scriptsize$\downarrow1.12$}\\
        Feature  w/o  Price Fluctuations &  1.65\textcolor{darkred}{\scriptsize$\downarrow1.06$} & 1.43 \textcolor{darkred}{\scriptsize$\downarrow0.80$} & 1.57 \textcolor{darkred}{\scriptsize$\downarrow1.00$}\\
        Feature  w/o  Influence from Other Regions &  1.79 \textcolor{darkred}{\scriptsize$\downarrow1.20$} & 2.15 \textcolor{darkred}{\scriptsize$\downarrow1.52$} & 1.70 \textcolor{darkred}{\scriptsize$\downarrow1.16$}\\
        Feature  w/o  Confidence Score &  2.13 \textcolor{darkred}{\scriptsize$\downarrow1.54$} & 2.28 \textcolor{darkred}{\scriptsize$\downarrow1.65$} & 2.02 \textcolor{darkred}{\scriptsize$\downarrow1.48$}\\ 
        \hline
        \textbf{FAEP (Ours)} &  \textbf{0.59} & \textbf{0.63} & \textbf{0.54} \\
        \hline
    \end{tabular}
    \label{tab:5}
    \vspace{-15pt}
\end{table*}

\subsubsection{Ablation of the core components:}
We assess the performance of two key components: LLM-based Feature Augmentation and Selection and the Hybrid Model, which integrates LSTM and XGBoost. The results of the ablation study are shown in Table \ref{tab:5}. Without LLM-augmented feature engineering, the model's MAE, MSE, and MAPE decrease by 1.04, 1.16, and 1.06, respectively. This highlights the effectiveness of the features enhanced by LLM. In the FAEP hybrid structure, if LSTM is removed, the model performance drops by 0.18 (MAE), 0.68 (MSE), and 0.19 (MAPE). Similarly, without XGBoost, the model’s performance decreases by 0.76 (MAE), 0.94 (MSE), and 0.74 (MAPE). This demonstrates that the LSTM and XGBoost components in the hybrid structure complement each other effectively, leading to better market prediction results.
\vspace{-10pt} 
\subsubsection{Ablation of feature selection:}
To systematically assess the impact of feature selection on model performance, we analyze four key categories: weather factors (Air Temperature, Wind Speed, Relative Humidity, Mean Sea-Level Pressure), market supply and demand (supply, demand, retail prices), price fluctuations ($CV_t$, $J_t$), and regional influences.

In Table \ref{tab:5}, the result demonstrates that each feature category significantly contributes to model performance. Removing any key feature, such as weather factors, market supply and demand, price fluctuations, or regional influences, results in notable declines in MAE, MSE, and MAPE, indicating that these features are essential for accurate market prediction. The model achieves the best performance when all features are included, highlighting the importance of comprehensive feature selection in improving prediction accuracy.
\vspace{-10pt}
\section{Conclusions}
\vspace{-5pt}
In this study, we predict electricity market price fluctuations in New South Wales, Australia. FAEP uses feature engineering and enhanced LLM with RAG to improve model accuracy. Key factors like weather, volatility jumps, and cross-market influences are incorporated, with precise weather feature extraction. Based on these features, we designed a hybrid XGBoost-LSTM model, boosting predictive performance. This combination of advanced feature engineering and hybrid machine learning effectively tackles electricity price forecasting challenges, demonstrating potential for broader applications.

%
%
\bibliographystyle{splncs04}
\bibliography{icic}

\end{document}